\documentclass[journal=jctcce,manuscript=article]{achemso}
\setkeys{acs}{articletitle = false}
\setkeys{acs}{super = false}

\usepackage[T1]{fontenc}
\usepackage[czech]{babel}
\usepackage[latin2]{inputenc}
\usepackage{graphicx}
\usepackage{amsmath}
\usepackage{color}
\usepackage{comment}
\usepackage{hyperref}
\usepackage{physics}

\title{Silně korelované systémy a metoda renormalizační grupy matice hustoty v kvantové chemii}

\author{Libor Veis}
\email{libor.veis@jh-inst.cas.cz}
\affiliation{Ústav fyzikální chemie Jaroslava Heyrovského, Akademie věd České republiky, v.v.i., Dolejškova 3, 18223 Praha 8}

\author{Jan Brandejs}
\affiliation{Ústav fyzikální chemie Jaroslava Heyrovského, Akademie věd České republiky, v.v.i., Dolejškova 3, 18223 Praha 8}

\author{Jiří Pittner}
\email{jiri.pittner@jh-inst.cas.cz}
\affiliation{Ústav fyzikální chemie Jaroslava Heyrovského, Akademie věd České republiky, v.v.i., Dolejškova 3, 18223 Praha 8}


\begin{document}

\textit{Rádi bychom tento text věnovali Prof. Rudolfovi Zahradníkovi k jeho devadesátým narozeninám s poděkováním za jeho celoživotní práci v oboru kvantové chemie inspirující generace kvantových chemiků, včetně nás samotných.}

\begin{abstract}
Tento článek přináší úvodní přehled metody renormalizační grupy matice hustoty a jejího použití v kvantové chemii.
Nejprve je zaveden pojem silně (staticky) korelovaných systémů a je vysvětleno, proč pro takové systémy tradiční metody kvantové chemie
nejsou vhodné. Poté je na základě motivace principem lokality zavedena aproximace vlnové funkce ve formě produktu matic (MPS)
a její grafické znázornění a způsob výpočtu středních hodnot operátorů. 
Další sekce zavádí kanonickou formu MPS rozvoje a vysvětluje její souvislost s algoritmem renormalizační grupy matice hustoty (DMRG), jako způsobem optimalizace matic v MPS rozvoji směrem k variační minimalizaci energie.
Následuje stručný přehled vlastností DMRG metody, je uvedena její výpočetní složitost a z ní vyplývající omezení
a je podán stručný přehled metod pro zahrnutí zvývající dynamické korelace. Závěrem je metoda ilustrována
na několika příkladech výpočtů reálných, chemicky netriviálních molekul.
\end{abstract}

\maketitle

\section{Úvod}

V nerelativistické teorii elektronové struktury molekul je centrální úlohou řešení Schr\"{o}dingerovy rovnice
popisující stav elektronů při pevných polohách atomových jader Hamiltoniánem
\begin{equation}
\label{Htotal}
\hat{H}_{\rm el} =
\underbrace{\sum_{A<B}^{\rm nucl.} {Z_A Z_B\over r_{AB}} }_{V_{\rm nn}}
\underbrace{-{1\over 2} \sum_i  \nabla_i^2}_{\hat{T}_{\rm e}}
\underbrace{-\sum_i\sum_A^{\rm nucl.} {Z_A \over r_{iA}} }_{V_{\rm en}}
+\underbrace{\sum_{i<j} {1\over r_{ij}}}_{V_{\rm ee}}
\end{equation}
Mezielektronová Coulombická interakce popsaná členem $V_{\rm ee}$ mimořádně komplikuje řešení této rovnice, což
vystihl Paul Dirac ve svém známém citátu: 
``Fundamentální zákony potřebné pro matematický popis velké části fyziky a celé chemie jsou tedy kompletně známé
a problém je jen v tom, že aplikace těchto zákonů vede na rovnice, které jsou příliš složité na to, aby mohly být vyřešeny.
''
Skutečně, nebýt členu $V_{ee}$, nalezení základního stavu Hamiltoniánu (\ref{Htotal}) by byla úloha téměř triviální,
neboť přesná vlnová funkce by byla antisymetrizovaným produktem (determinantem) jednočásticových funkcí (spinorbitalů).
Korelační energií rozumíme rozdíl mezi energií přesného řešení (\ref{Htotal}) a přibližným řešením Hartree-Fockovou (HF) metodou,
která používá právě tento typ přibližné vlnové funkce; obecněji vliv  členu $V_{ee}$ nazýváme korelačními efekty.

Protože korelační energie tvoří jen asi jedno procento energie celkové, byla 
v raném věku kvantové chemie vkládána velká naděje v HF metodu, která
nahrazuje mezielektronovou repulzi jednočásticovým operátorem středního selfkonzistentního pole, což koresponduje
s jednodeterminantální vlnovou funkcí. 
Bohužel se brzy ukázalo, že tato aproximace je pro chemii příliš nepřesná, neboť pro kvantifikaci chemické reaktivity
jsou klíčové energetické rozdíly, velmi malé oproti celkové energii a právě příspěvek
korelační energie do těchto rozdílů hraje významnou roli. 
Kompenzaci chyb způsobených zanedbáním korelační energie lze očekávat pouze u třídy tzv. homodesmických reakcí \cite{carsky_urban},
kam ale naprostá většina chemických reakcí nespadá.
Naproti tomu např. vliv relativitických efektů na energie vnitřních slupek, které se přímo neúčastní
chemických vazeb, se u lehčích atomů do velké míry kompenzuje a dá se tedy zanedbat, 
přestože v absolutní velikosti může být větší než korelační energie.
Je tedy jasné, že snaha o zahrnutí korelační energie co možno nejpřesněji při realistické výpočetní náročnosti je
motivací pro vývoj stále lepších výpočetních metod.

Ve výpočetní praxi se vždy problém Hamiltoniánu (\ref{Htotal}) diskretizuje pomocí konečného počtu bázových funkcí (pro molekuly nejčastěji
báze Gaussovských atomových orbitalů) a výpočtem typu self-konzistentního pole se převede do báze ortonormálních
molekulových spinorbitalů.
Elektronový Hamiltonián  (\ref{Htotal})  pak ve formalizmu druhého kvantování lze zapsat ve tvaru
\begin{equation}
\label{H2q}
\hat{H} = V_{\rm nn} + \sum_{pq=1}^k h_{pq} a_p^\dagger a_q + \sum_{pqrs=1}^k v_{pqrs} a_p^\dagger a_q^\dagger a_s a_r ,
\end{equation}
kde $h_{pq}$ a $v_{pqrs}$ jsou jedno- a dvou-elektronové integrály.
Tato forma Hamiltoniánu je v jistém směru dokonce obecnější než (\ref{Htotal}), neboť zahrnuje i relativistickou kvantovou chemii (s patřičně zobecněnými  $h_{pq}$ a $v_{pqrs}$),
pokud zanedbáme efekty kvantové elektrodynamiky, např. tvorbu virtuálních elektron-pozitronových párů.

Proč je nalezení základního stavu (a ev. excitovaných stavů) Hamiltoniánu (\ref{H2q}) tak obtížné, že ani půlstoletí
úsilí fyziků a kvantových chemiků nepřineslo definitivní řešení a grantové agentury na celém světě dále věnují peníze daňových poplatníků
na vývoj nových výpočetních metod?
Rigorózní odpověď na tuto otázku přinesla až poměrně nedávno trojice matematiků Julia Kempe, Alexei Kitaev a Oded Regev 
v práci \cite{kempe2006}, kde ukázali, že problém základního stavu Hamiltoniánu (\ref{H2q}) patří do kategorie tzv. NP-úplných problémů.
To znamená, že vhodnou volbou koeficientů $h_{pq}$ a $v_{pqrs}$ je v principu možné do základního stavu Hamiltoniánu  (\ref{H2q}) zakódovat řešení jiného matematického  NP-úplného problému, např. známého problému obchodního cestujícího z teorie grafů.
V tomto článku nemůžeme zabíhat do teorie výpočetní složitosti, takže jen zmíníme, že pro NP-úplné problémy nejsou známy
obecné a výpočetně efektivní (tzn. škálující polynomiálně s velikostí systému, na rozdíl od řešení hrubou silou, škálujícího exponencielně) algoritmy na jejich řešení (a většina matematiků se domnívá, že ani neexistují, což se však zatím nepodařilo dokázat).
Aby kvantová chemie dokázala vyvinout efektivní metody na řešení problému (\ref{H2q}), musí to tedy být na úkor obecnosti, tj. musí se vytvářet vhodné aproximace,
které jsou použitelné pro fyzikálně relevantní Hamiltoniány tohoto typu (a nebudou tedy obecně fungovat pro libovolně zvolené  hodnoty $h_{pq}$ a $v_{pqrs}$).
To je velmi náročné na invenci, o čemž svědčí i udělení Nobelovy ceny za chemii Walteru Kohnovi a Johnu Poplovi v roce 1998 za vývoj metody funkcionálu hustoty \cite{dft_book} resp. dalších výpočetních metod kvantové chemie
a  Nobelova cena za fyziku, udělená roku 1982  Kennethu Wilsonovi
za metodu numerické renormalizační grupy. V roce 1992 pak Steve White na základě Wilsonovy metody vytvořil
metodu renormalizační grupy matice hustoty (DMRG) \cite{white_1992, white_1993}, jíž se v tomto článku budeme zabývat.

Přestože výpočet přesného základního stavu (\ref{H2q}) je tak obtížný, formálně tento stav můžeme snadno vyjádřit ve formě
normalizované lineární kombinace Slaterových determinantů, které lze vytvořit pro $N$ elektronů v $k$ orbitalech,
resp. s omezením na $N_\alpha$ elektronů  se spinem $1/2$ a  $N_\beta$ elektronů  se spinem $-1/2$  v $k$ orbitalech, neboť $\hat{H}_{\rm el}$ komutuje s $\hat{S}_z$.
Rozvoj můžeme vytvořit tak, že vyjdeme z referenčního HF determinantu $|\Phi_0\rangle$ a vytvoříme všechny mono-, bi-, tri-, a vyšší excitace
\begin{equation}
\label{fci1}
|\Psi\rangle = C_0  |\Phi_0\rangle + \sum_{ia} C_i^a  |\Phi_i^a\rangle + \sum_{i<j, a<b} C_{ij}^{ab} |\Phi_{ij}^{ab}\rangle + \cdots = \sum_I C_I |\Phi_I\rangle .
\end{equation} 
Aproximace tohoto rozvoje dle excitační úrovně je základem klasických kvanově chemických metod a dobře funguje, pokud má HF determinant dominantní váhu.
Rozvoj (\ref{fci1}) ale můžeme uvažovat i bez volby referenčního determinantu jako lineární kombinaci všech možných Slaterových determinantů pro daný počet elektronů  a orbitalů,
z čehož ihned plyne počet jeho členů  (tj. dimenze příslušného Hilbertova prostoru)
\begin{equation}
\label{dimH}
{\rm dim} \; {\cal H}(N_\alpha, N_\beta, k) = {k \choose N_\alpha}{k \choose N_\beta} .
\end{equation}
Rozvoj (\ref{fci1})  také dovoluje zavést termíny statická (silná) a dynamická (slabá) korelace.
Pokud koeficient u HF determinantu $C_0$ je mnohem vyšší než ostatní koeficienty, mluvíme o dynamické korelaci.
V opačném případě, pokud nelze nalézt jediný dominantní koeficient, tzn. existuje skupina podobně velkých nejvyšších koeficientů (v absolutní hodnotě), jde o silně (staticky) korelovaný systém, alternativně nazývaný také multireferenční.
K této situaci dochází typicky  tehdy, když několik molekulových orbitalů na rozhraní mezi obsazenými a neobsazenými
v HF determinantu má podobnou energii, což nastává např. při disociaci vazeb, u diradikálů, u sloučenin přechodných kovů aj.
	\footnote{Větší počet dominantních determinantů může mít vlnová funkce i z důvodu spinové adaptace pro stav s $|M_s| < S$; toto není skutečný případ silně korelovaného systému, ale z praktického hlediska též vyžaduje multireferenční přístup, a proto jej nebudeme uvažovat zvlášť.}
Statická korelace je tedy přítomna jen někdy, ale dynamická korelace je přítomna vždy, což odpovídá obrovskému množství členů s malým koeficientem v rozvoji (\ref{fci1}), které ale v souhrnu nejsou vůbec zanedbatelné.
Zmíněnou skupinu orbitalů důležitých pro statickou korelaci budeme dále nazývat aktivním prostorem.
Poznamenejme též, že přechod mezi statickou a dynamickou korelací není ostrý, aktivní prostor MO lze v principu stále rozšiřovat,
až do limitu všech MO a postupně tedy přidávat stále více dynamické korelace.

Kvantově chemické metody vhodné pro pouze dynamicky korelované systémy jsou již velmi dobře probádány a jsou rutinně aplikovány
na mnoho nejrůznějších systémů - jde zejména o metody funkcionálu hustoty (DFT)\footnote{DFT s exaktním Hohenberg-Kohnovým funkionálem musí dát přesnou energii i pro silně korelované systémy, ale v současnosti dostupné aproximativní funkcionály nefungují v tomto případě spolehlivě.}, 
konfigurační interakci (\textit{configuration interaction}, CC) s omezením excitačního ranku,
 poruchové metody M{\o}llerova-Plessetova typu a metody spřažených klastrů (\textit{coupled clusters}, CC), např. CCSD(T) \cite{all-encyclop}.
Jejich další vývoj směřuje zejména ke zpřesňování DFT funkcionálů a ke snižování výpočetního škálování směrem k lineární závislosti
na velikosti systému s využitím lokality (např. DLPNO-CC metoda \cite{dlpno}).

Naprosto odlišná situace platí pro systémy se silně korelovanými elektrony.
Z teoretického hlediska je hlavním problémem silné korelace velký počet determinantů, které významně přispívají do vlnové funkce. Navíc s přibývajícím počtem silně korelovaných orbitalů jejich počet roste \textit{exponenciálně}. Mohlo by se tedy zdát, že problém silné korelace není klasicky řešitelný.
Tradičně se v kvantové chemii pro popis těchto systémů
používá metoda CASSCF a CASPT2 \cite{roos_book}, multireferenční CI \cite{szalay_review},
mnoho úsilí bylo věnováno i multireferenčnímu zobecnění CC metod \cite{ourspringer, lyakh_review}.
Žádná z těchto metod však neřeší problém exponenciální složitosti a jsou tedy použitelné jen pro malé aktivní prostory.
V poslední době se objevila řada prací stochasticky samplujících rozvoj (\ref{fci1}) metodou Quantum Monte Carlo dosahujících vynikající přesnosti \cite{booth_2009}.
Metoda DMRG \cite{white_1992, white_1993} je další způsob výpočetně efektivního nalezení téměř přesného rozvoje  (\ref{fci1}), původně vyvinutý pro modely fyziky pevné fáze (např. Hubbardův model), a později převzatý a dále rozvinutý pro kvantově chemický Hamiltonián  (\ref{H2q}).
Předem podotkněme, že jak FCIQMC tak i DMRG metody nejsou schopné zahrnout všechny molekulové orbitaly a popsat tak dynamickou korelaci,
mohou však oproti tradičnímu CASSCF být použity na mnohem větší aktivní prostory.

Na závěr první kapitoly bychom rádi poznamenali, že tento text je pouze lehkým úvodem do teorie renormalizačni grupy matice hustoty v kvantové chemii inspirovaným
vynikajícím přehledným článkem Chana a Sharmy \cite{chan_review}. Prakticky výhradně se věnujeme modernímu pojetí metody DMRG, které je založené na specifickém tvaru vlnové funkce a jejích vlastnostech. Zvídavé čtenáře bychom rádi odkázali na detailní přehledné články, kterých je k dispozici nespočet, např. \cite{schollweck_2005, schollwock_2011, legeza_review}.

\section{Princip lokality}

Problém principiálně exponenciální složitosti silné korelace nastíněný v předešlé kapitole je však z velké části umělý. Přestože kvantová mechanika v principu připouští existenci různých exotických stavů, jako například stavy slavné Schr\"{o}dingerovy kočky, většina Hilbertova prostoru není podle naší zkušenosti okupována v přírodě se vyskytujícími základními a nízko ležícími excitovanými stavy. Tyto stavy zaujímají pouze maličkatou část tohoto prostoru, často nazývanou ``přirozený koutek Hilbertova prostoru'' \cite{eisert_tensor_networks}.

Opomineme-li takové problémy, jako jsou např. fázové přechody kritických mnohačás\-ti\-co\-vých kvantových systémů a zaměříme se výhradně na chemii, jeden z nejdůležitějších v chemii platných empirických zákonů je princip lokality. Bez ohledu na složitost studované molekuly, její odezva na externí poruchu vždy zůstává lokální.
Jinými slovy, reakce na jedné části molekuly nebo materiálu nezpůsobí okamžité změny v části, která je makroskopicky vzdálená.

Přestože vlnová funkce silně korelovaného stavu může být rozvojem \textit{exponenciálního} množství determinantů, koeficienty těchto determinantů by měly být vysoce strukturované, aby odrážely princip lokality. Ve skutečnosti by taková vlnová funkce měla být parametrizovatelná počtem parametrů proporcionálním velikosti studovaného systému. Otázkou tedy je, jak zjednodušit (zhustit) vlnovou funkci silně korelovných stavů zavedením lokality. V následující kapitole ukážeme, že metoda DMRG \cite{white_1992, white_1993}, nám tuto možnost poskytuje.

\section{Vlnová funkce metody renormalizační grupy matice hustoty}
\label{section_wf}

Metoda DMRG pochází z fyziky pevné fáze, kde byla původně vyvinuta pro účely studia jednodimenzionálních kvantových systémů s krátkodosahovou interakcí \cite{white_1992, white_1993}. 
Její úspěchy ve fyzice pevné fáze \cite{schollweck_2005} byly mimo jiné motivací pro aplikaci v kvantové chemii \cite{white_1999}. Kvantově chemická verze metody DMRG byla v průběhu předešlých téměř dvaceti let úspěšně použita pro výpočty elektronové struktury celé řady molekul \cite{chan_2002, legeza_2003a, chan_review, legeza_review, wouters_review, yanai_review, marti_2010} (více o aplikaci v kapitole \ref{section_applications}) a má díky svým vlastnostem diskutovaným dále bez pochyb nezastupitelné místo mezi výpočetními metodami pro molekuly se silně korelovanými elektrony.

Pro demonstraci speciálního tvaru DMRG vlnové funkce, uvažujme nyní přesný (FCI) rozvoj vlnové funkce (\ref{fci1}) rozepsaný trochu jiným způsobem

\begin{equation}
  \ket{\Psi} = \sum_{\{n\}} C^{n_1 n_2 \ldots n_k} \ket{n_1 n_2 \ldots n_k}.
  \label{eq_fci}
\end{equation}

\noindent
Jednotlivé determinanty $\ket{n_1 n_2 \ldots n_k}$ nejsou charakterizovány danou excitací vůči referenčnímu HF determinantu jako v rozvoji (\ref{fci1}), nýbrž obsazeností všech orbitalů. Báze každého z celkem $k$ orbitalů odpovídá

\begin{equation}
 n_i \in \{ \ket{~}, \ket{\uparrow}, \ket{\downarrow}, \ket{\uparrow \downarrow} \},
 \label{basis}
\end{equation}

\noindent
přičemž obsazovací čísla jednotlivých orbitalů se mohou libovolně kombinovat.
Rozvojové koeficienty CI z rovnice (\ref{fci1}) jsou přeorganizovány do formy tenzoru $C^{n_1 n_2 \ldots n_k}$. Takto reprezentované determinanty tedy mohou obsahovat libovolný počet elektronů dovolený Pauliho vylučovacím principem, od vakua až po všechny orbitaly plně zaplněné\footnote{Jedná se o reprezentaci v tzv. Fockově prostoru.}. 
Vlnová funkce $\ket{\Psi}$ však musí splňovat správný počet $\alpha$ a $\beta$ elektronů, což znamená, že elementy $C^{n_1 n_2 \ldots n_k}$ odpovídající jinému počtu elektronů, než $N_\alpha, N_\beta$ budou nulové.
Správná antisymetrie $\ket{\Psi}$ je zajištěna antikomutačními relacemi kreačních a anihilačních operátorů, resp. jejich maticové reprezentace.

Pro úplnost ještě uveďme maticové reprezentace kreačních operátorů\footnote{Matice anihilačních oprátorů lze získat Hermitovským sdružením matic kreačních operátorů.} v bázi jednoho orbitalu (\ref{basis}), které jsou potřebné při konstrukci Hamiltoniánu v kapitole \ref{section_algorithm}

\begin{equation}
a^{\dagger}_{\uparrow} = \begin{pmatrix} 0 & 0 & 0 & 0 \\ 1 & 0 & 0 & 0 \\ 0 & 0 & 0 & 0 \\ 0 & 0 & -1 & 0 \end{pmatrix} , \qquad a^{\dagger}_{\downarrow} = \begin{pmatrix} 0 & 0 & 0 & 0 \\ 0 & 0 & 0 & 0 \\ 1 & 0 & 0 & 0 \\ 0 & 1 & 0 & 0 \end{pmatrix}.
\end{equation}

\noindent
Zvídavý čtenář si snadno sám ověří správnou transformaci bázových funkcí (\ref{basis}) a také platnost antikomutačních relací.

Je zřejmé, že problém přesného rozvoje (\ref{eq_fci}) spočívá ve faktu,
že dimenze tenzoru rozvojových koeficientů $C$ roste s velikostí systému jako $4^k$, což je dokonce víc než velikost Hilbertova prostoru v rovnici (\ref{dimH}).
Navíc v případě silně korelovaných stavů je velký počet těchto koeficientů nenulový a nelze tak automaticky využít řídkosti $C$ jako v případě metod CI, či CC. Ukazuje se však, že pro mnoho typů silně korelovaných systémů (včetně těch relevantních pro kvantovou chemii) se tento tenzor dá velmi přesně aproximovat s využitím lokality.

Nejjednodušší možností jak zavést do rovnice (\ref{eq_fci}) lokalitu, by bylo aproximovat vysokodimenzionální tenzor rozvojových koeficientů $C$ (tenzor $k$-tého řádu) jako tenzorový součin vektorů okupancí (dimenze 4), tedy pro danou okupanci $\{ n_1, n_2 \ldots n_k \}$ jako součin jejich prvků (skalárů)

\begin{equation}
  C^{n_1 n_2 \ldots n_k} \approx A[1]^{n_1} A[2]^{n_2} \cdots A[k]^{n_k}. 
  \label{eq_scalar}
\end{equation}

\noindent
Číslo v hranatých závorkách značí, že prvky vektorů $A^{n}$ jsou specifické pro daný orbital.
Tento rozvoj sníží počet variačních parametrů z původních $4^k$ na $4k$, nicméně není v obecných případech pro svou malou flexibilitu dostatečně přesný.

Pro zvýšení počtu variačních parametrů a flexibility rozvoje v rovnici (\ref{eq_scalar}) můžeme nahradit prvky vektorů (skaláry) maticemi

\begin{equation}
  A[j]^{n_j} \rightarrow A[j]^{n_j}_{i i^{\prime}}.
\end{equation}

\noindent
Zavedeme tedy nové pomocné indexy $i$ a $i^{\prime}$. Aby tyto indexy nefigurovaly ve výsledném tenzoru rozvojových koeficientů $C$, musí přes ně probíhat sumace, jsou tzv. kontrahovány. Výsledkem je, že se rozvojové koeficienty $C$ aproximují jako

\begin{equation}
  C^{n_1 n_2 \ldots n_k} \approx \sum_{i_1 i_2 \ldots i_{k-1}} A[1]^{n_1}_{i_1} A[2]^{n_2}_{i_1 i_2} A[3]^{n_3}_{i_2 i_3} \cdots A[k]^{n_k}_{i_{k-1}}.
  \label{eq_mps_ind}
\end{equation}

\begin{figure}[!ht]
  \begin{minipage}{0.65\textwidth}
    \includegraphics[width=8.3cm]{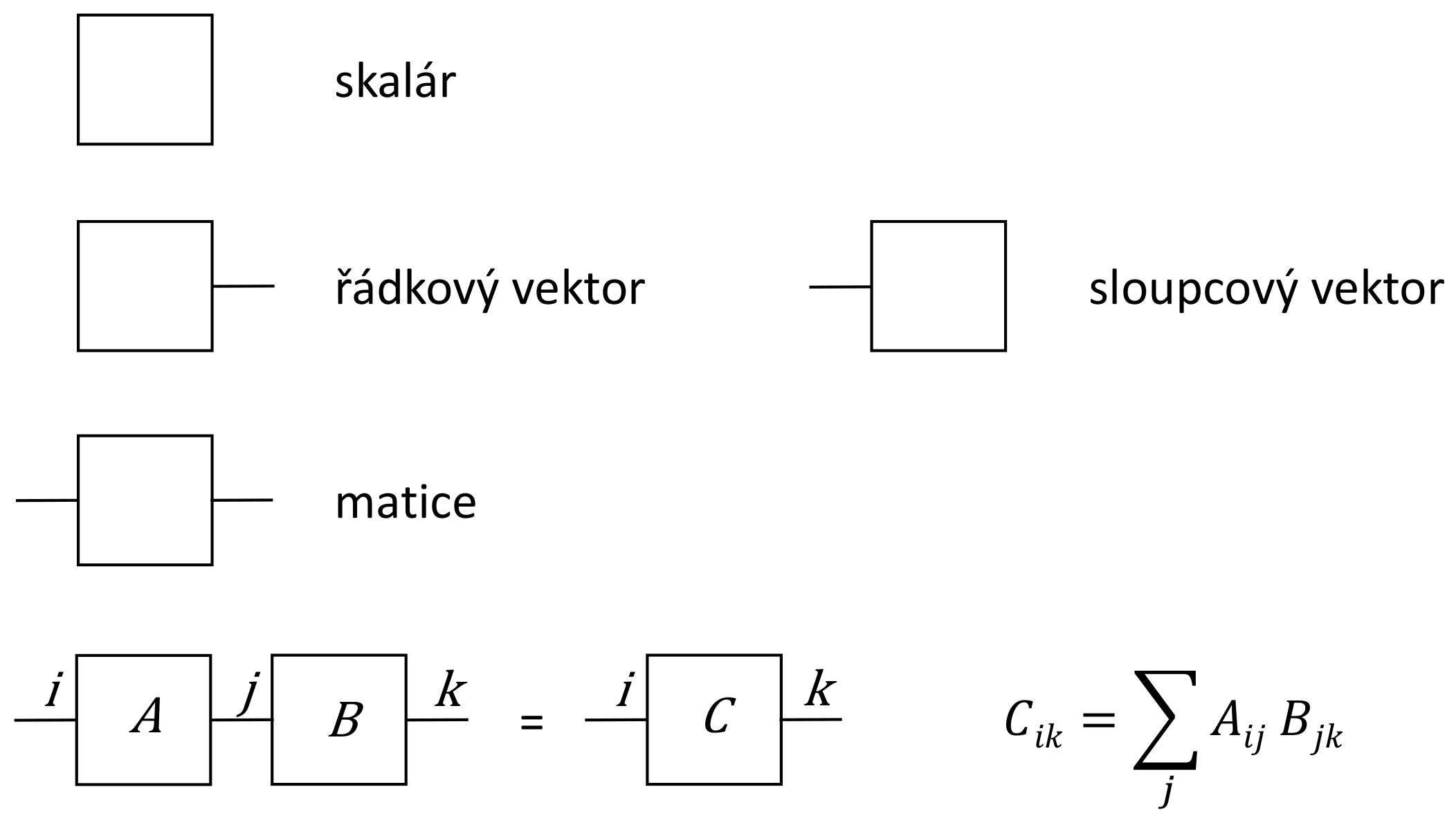}
    \begin{center} (a) \end{center}
    \vskip 1cm  
    \includegraphics[width=10cm]{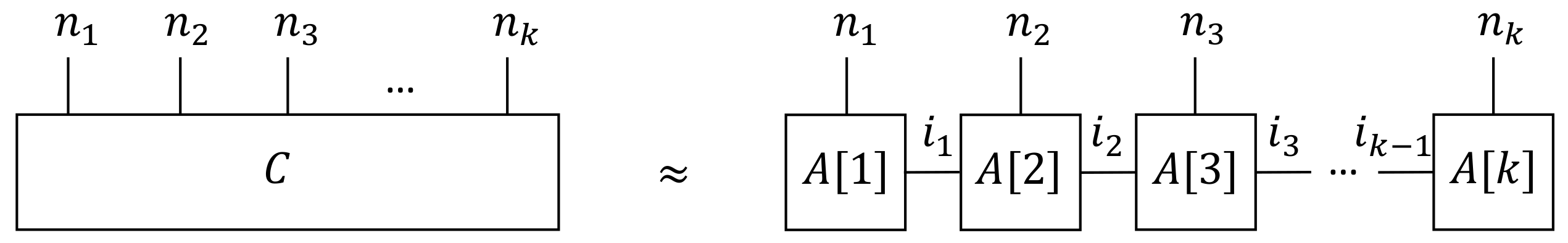}
    \begin{center} (b) \end{center}
  \end{minipage}    
  \begin{minipage}{0.3\textwidth}
    \includegraphics[width=5.1cm]{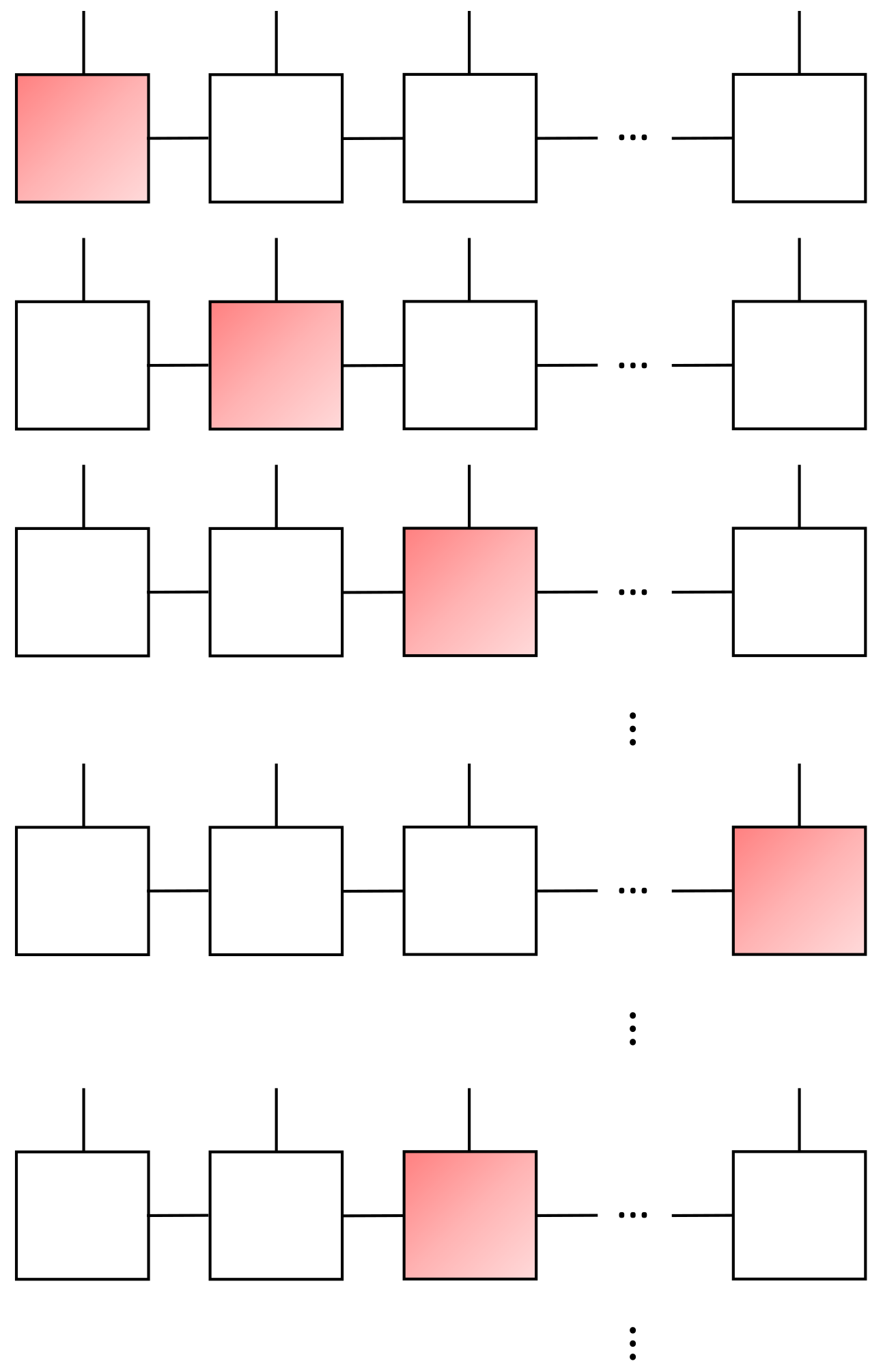}
    \begin{center} (c) \end{center}    
  \end{minipage}    
  \caption{Grafické tenzorové značení. (a) Tenzor $k$-tého řádu je reprezentován čtvercem s $k$ vnějšími čarami, které odpovídají jednotlivým indexům (skalár s žádnou, vektor s jednou, matice se dvěma, atd.). Pokud čára indexu spojuje dva tenzory, probíhá přes tento index sumace (tzv. kontrakce), jak je ukázáno na příkladu maticového násobení. (b) Aproximace tenzoru rozvojového koeficientu přesné vlnové funkce ($k$-tého řádu) pomocí MPS. (c) Iterativní procedura optimalizace MPS pomocí metody DMRG, tzv. ``zametání''. V dané iteraci se optimalizuje pouze jedna MPS matice.}
  \label{tensors}
\end{figure}

\noindent
Velmi užitečným nástrojem při práci s tenzory je jejich grafické značení stručně představené na obrázku \ref{tensors}a. Standardně se nově zavedené pomocné indexy, často označované jako tzv. virtuální indexy, značí vodorovnými čarami, naopak indexy odpovídající okupanci, tzv. fyzikální indexy, svislými čarami. Rovnice (\ref{eq_mps_ind}) je pak v tomto značení vyjádřena na obrázku \ref{tensors}b. Z důvodu přehlednosti nebudeme v dalším textu zdůrazňovat příslušnost matic $A^{n}_{i i^{\prime}}$ jednotlivým orbitalům pomocí čísla v hranatých závorkách. Pro kompaktnější vyjádření lze také využít maticového zápisu

\begin{equation}
  C^{n_1 n_2 \ldots n_k} \approx \mathbf{A}^{n_1} \mathbf{A}^{n_2} \mathbf{A}^{n_3} \cdots \mathbf{A}^{n_k}.
  \label{eq_mps}
\end{equation}

Výsledné rozvojové koeficienty jsou ve skutečnosti aproximovány jako součin matic $\mathbf{A}^{n}$, odtud také pramení název tohoto rozvoje, tzv. \textit{matrix product states} (MPS) \cite{schollwock_2011}. 
Dvě poznámky stojí za zmínku. Tou první je, že se v případě prvního a posledního orbitalu nejedná o matice, nýbrž o vektory. V případě prvního orbitalu a řádkový a v případě posledního o sloupcový. Jedině tak může být výsledkem součinu skalár výsledného koeficientu. Druhou poznámkou pak je, že se obecně nemusí jednat o aproximaci. Libovolnou vlnovou funkci lze ve skutečnosti \textit{přesně} vyjádřit v MPS tvaru \cite{vidal_2003}. V takovém případě však potřebný rozměr MPS matic obecně škáluje s původní dimenzí tenzoru rozvojových koeficientů $C$. Aby se jednalo o \textit{efektivní} faktorizaci, je potřeba jejich rozměr shora omezit. Pro jednoduchost předpokládejme, že rozměr všech MPS matic je stejný, roven $M$. Konečný počet variačních parametrů v takovém případě odpovídá $\mathcal{O}(4M^2k)$.

Přestože to nebylo zřejmé z původní formulace metody DMRG \cite{white_1992, white_1993}, rozvoj v rovnici (\ref{eq_mps}) odpovídá ve skutečnosti DMRG vlnové funkci \cite{ostlund_1995}. V průběhu DMRG algoritmu se variačně optimalizují prvky MPS matic, $A^{n}_{i i^{\prime}}$. Jedná se o iterativní proceduru (s jistou analogií k HF metodě self-konzistentního pole \cite{chan_2008}), v dané iteraci se optimalizují prvky pouze jedné MPS matice (ostatní zůstávájí fixní), jak je naznačeno na obrázku \ref{tensors}c.
Se zvětšováním dimenze $M$ můžeme libovolně zvyšovat přesnost metody DMRG\footnote{Jedná se o variační metodu, takže větší počet variačních parametrů musí zákonitě odpovídat přesnější energii.}, až eventuelně získat přesnou energii. 

Zkombinováním rovnic (\ref{eq_mps}) a (\ref{eq_fci}) získáme finální předpis pro DMRG vlnovou funkci

\begin{equation}
  \ket{\Psi_{\text{DMRG}}} = \sum_{\{n\}} \mathbf{A}^{n_1} \mathbf{A}^{n_2} \mathbf{A}^{n_3} \cdots \mathbf{A}^{n_k} \ket{n_1 n_2 \ldots n_k}.
  \label{eq_dmrg_wf}
\end{equation}

\subsection{Lokalita v renormalizační grupě matice hustoty}

Jak již bylo zmíněno, princip lokality hraje naprosto klíčovou roli pro efektivní popis silně korelovaných systémů. DMRG vlnová funkce v sobě ve skutečnosti zahrnuje lokalitu pro případ jendodimenzionálních systémů. To je také důvod velké přesnosti metody DMRG při popisu jednodimenzionálních problémů ve fyzice pevné fáze \cite{schollweck_2005}.

Abychom si osvětlili, jak konkrétně je lokalita v DMRG vlnové funkci zahrnuta, uvažujme pro jednoduchost např. jednodimenzionální řetízek vodíkových atomů v minimální bázi (jeden $s$  orbital na jeden atom vodíku). Při pohledu na rovnici (\ref{eq_mps_ind}) vidíme, že index $i_1$ je spjatý s okupancemi $n_1$ a $n_2$ a prvními dvěma MPS maticemi (lépe řečeno vektorem a maticí), $A[1]^{n_1}_{i_1}$ a $A[2]^{n_2}_{i_1 i_2}$. Pokud by rovnice (\ref{eq_mps_ind}) neobsahovala sumaci (kontrakci) přes index $i_1$, pak by se vlnová funkce dala napsat jako součin části odpovídající prvnímu orbitalu ($n_1$) a části odpovídající zbylým orbitalům\footnote{Taková situace by odpovídala neexistenci interakce mezi prvním atomem vodíku a zbylými atomy.}. Index $i_1$ a jeho kontrakce zákonitě umožňují korelaci mezi obsazenostmi $n_1$ a $n_2$. Obdobně index $i_2$ zprostředkovává korelaci mezi obsazenostmi $n_2$ a $n_3$ a stejně pro ostatní pomocné indexy. Přestože není žádné přímé spojení mezi ne-sousedními orbitaly přes pomocné indexy (jako např. $n_1$ a $n_3$), neznamená to, že by okupance $n_1$ a $n_3$ nebyly korelované. Korelace mezi nimi je zprostředkovaná skrz korelaci $n_1 \leftrightarrow n_2$ a $n_2 \leftrightarrow n_3$.

Pokud pracujeme s lokálními orbitaly a topologie molekuly je lineární, nebo alespoň blízká lineární (kvazi-lineární), pak zmíněná sekvenční struktura korelací přesně odpovídá realitě. 
Ve skutečnosti je MPS vlnová funkce optimálním popisem jednodimenzionálních problémů s krátkodosahovou interakcí, což se dá rigorózně dokázat a je ve vztahu k tzv. \textit{area law} kvantové informatiky \cite{eisert_2010}.
Formálně tedy můžeme vidět metodu DMRG jako řešení problému silné korelace v jedné dimenzi.

Otázkou zůstává, jak je tomu u nelineárních molekul. Obecně lze říci, že v případě více-dimenzionálních problémů (obecných molekul) metoda DMRG \textit{nezahrnuje} lokalitu optimálním způsobem a složitost problému silné korelace zůstává  \textit{exponenciální}. Metoda DMRG je v principu schopná odstranit toto škálování jen pro jednu dimenzi. Efektivní popis obecných molekul vyžaduje více flexibilní vlnovou funkci. Velkým příslibem v tomto směru je nedávný vývoj v oblasti tenzorových sítí \cite{verstraete_2008}, konkrétně stromových tenzorových sítí \cite{nakatani_2013, murg_2014, gunst_2018}.

Nicméně i v případě obecných silně korelovaných molekul je metoda DMRG velmi úspěšná a při správném řazení orbitalů v jednodimenzionálním uspořádání (více v kapitole \ref{section_algorithm}) je schopná pracovat s aktivními prostory výrazně většími nežli umožnujě např. metoda CASSCF.

\subsection{Kanonický tvar DMRG vlnové funkce}
\label{canonical_form}

Pozornému čtenáři jistě neunikne, že tvar DMRG vlnové funkce v rovnici (\ref{eq_dmrg_wf}) není jednoznačný. Ve skutečnosti můžeme mezi jakékoliv dvě MPS matice vložit identitu $\mathbf{X}\mathbf{X}^{-1}$, kde $\mathbf{X}$ je libovolná invertovatelná matice, a vlnová funkce se takto nezmění, je vůči této operaci invariantní. To je v jisté analogii k HF metodě, která je invariantní vůči rotaci obsazených orbitalů \cite{chan_2008}.

V principu by se daly prvky všech MPS matic optimalizovat přímo, např. pomocí Newto\-no\-vy-Raph\-so\-no\-vy techniky, avšak zmíněná nejednoznačnost by vedla k numerickým problémům. Jak již bylo zmíněno, v praxi se využívá iterativní algoritmus \cite{white_1992, white_1993}, který je velmi robustní a jeho výsledkem je vlnová funkce ve speciálním, tzv. kanonickém tvaru

\begin{equation}
  \ket{\Psi} = \sum_{\{ n \}} \mathbf{L}^{n_1} \cdots \mathbf{L}^{n_{p - 1}} \mathbf{C}^{n_p} \mathbf{R}^{n_{p + 1}} \cdots \mathbf{R}^{n_k} \ket{n_1 n_2 \ldots n_k},
  \label{eq_mps_can}
\end{equation}  
  
\noindent
kde matice $\mathbf{L}^{n}$ a $\mathbf{R}^{n}$ splňují následující podmínky ortonormality

\vskip -0.5cm
\begin{eqnarray}
  \sum_n \mathbf{L}^{n\dagger} \mathbf{L}^{n} & = & \mathbf{I} , \label{eq_ortonotmalita} \\
  \sum_n \mathbf{R}^{n} \mathbf{R}^{n \dagger} & = & \mathbf{I} . \label{eq_ortonotmalita2}
\end{eqnarray}
  
 \noindent
V dané iteraci se získají prvky matic $\mathbf{C}^n$, které minimalizují energii (zbylé MPS matice mají tvar získaný v předešlých iteracích) a přechod do další iterace je realizován rozkladem na singulární hodnoty (\textit{singular value decomposition}, SVD)\footnote{Abychom mohli využít rozkladu na singulární hodnoty, musíme nejprve přeorganizovat tenzor $C^n_{lr}$ do tvaru matice sloučením jednoho pomocného indexu s fyzikálním, např. $C^n_{lr} \rightarrow C_{(nl),r}$, kde $(nl)$ označuje nově vzniklý složený index.}. SVD představuje velmi užitečný nástroj lineární algebry, pomocí něhož lze reprezentovat libovolnou matici  $\mathbf{M}$ jako
  
\begin{equation}
  \mathbf{M} = \mathbf{U} \mathbf{D} \mathbf{V}^{\dagger},
\end{equation}
   
\noindent 
kde matice $\mathbf{U}$ je charakterictická tím, že má ortonormální sloupce (je tvořena tzv. levými singulárními vektory), zatímco matice $\mathbf{V}^{\dagger}$ obsahuje ortonormální řádky (tvořena pravými singulárními vektory) a $\mathbf{D}$ je diagonální matice s nezápornými hodnotami (tzv. singulární hodnoty). Při přechodu do další iterace směrem doprava (viz. obrázek \ref{tensors}c) se uplatní matice $\mathbf{U}$, naopak při přechodu doleva matice $\mathbf{V}^{\dagger}$ a z jejich zmíněných vlastností vyplývají podmínky ortonormality (\ref{eq_ortonotmalita}) a (\ref{eq_ortonotmalita2}).
   
Původní formulace metody DMRG \cite{white_1992, white_1993} není založená na MPS tvaru vlnové funkce, ale využívá aparátu renormalizační grupy. Podle našeho názoru není z pedagogického hlediska pro uvedení do problematiky velmi vhodná a proto se jí nebudeme detailně věnovat, raději odkážeme čtenáře na velmi obsažné přehledné články \cite{schollweck_2005, legeza_review}. Při formulaci renormalizační grupy odpovídá nastíněná procedura rozvoji vlnové funkce do báze tzv. levého bloku, $\{ l \}$, příslušného orbitalu, $\{ n_p \}$, a tzv. pravého bloku, $\{ r \}$

\begin{equation}
  \ket{\Psi} = \sum_{lnr} C_{lr}^{n} \ket{l n_p r}.
  \label{eq_blocks}
\end{equation}  

\noindent
Báze $\{ l \}$ je renormalizovaná \textit{mnoha-elektronová} báze odpovídající prostoru orbitalů $[1 \ldots (p-1)]$, obdobně báze $\{ r \}$ patří prostoru orbitalů $[(p+1) \ldots k]$. Pokud formálně vyjádříme renormalizované báze levého a pravého bloku pomocí matic $\mathbf{L}^n$ a $\mathbf{R}^n$
   
\vskip -0.5cm
\begin{eqnarray}
  \ket{l_i} & = & \sum_{\{n\}} [\mathbf{L}^{n_1} \mathbf{L}^{n_2} \cdots \mathbf{L}^{n_{p-1}}]_i \ket{n_1 n_2 \ldots n_{p-1}}, \\
  \ket{r_i} & = & \sum_{\{n\}} [\mathbf{R}^{n_{p+1}} \mathbf{R}^{n_{p+2}} \cdots \mathbf{R}^{n_k}]_i \ket{n_{p+1} n_{p+2} \ldots n_k} .
\end{eqnarray} 
 
\noindent
Vidíme, že determinantový rozvoj v rovnici (\ref{eq_mps_can}) je ekvivalentní rozvoji v menší ortonormální renormalizované bázi (\ref{eq_blocks}).
 
Protože platí, že singulární vektory matice $\textbf{M}$ odpovídají vlastním vektorům matic $\mathbf{M}\mathbf{M}^{\dagger}$ a $\mathbf{M}^{\dagger}\mathbf{M}$, je tvorba redukovaných matic hustoty levého, či pravého bloku zvětšených o $p$-tý orbital

\begin{equation}
  \rho^{\text{[1 \ldots p]}}_{ln, l^{\prime}n^{\prime}}= \sum_r C^n_{lr} C^{n^{\prime}}_{l^{\prime} r} \qquad \rho^{\text{[p \ldots k]}}_{nr, n^{\prime}r^{\prime}}= \sum_l C^n_{lr} C^{n^{\prime}}_{l r^{\prime}} 
\end{equation} 

\noindent
a jejich následná diagonalizace ekvivalentní k SVD. Odtud mimo jiné také pochází termín matice hustoty v názvu DMRG metody.

Na závěr této kapitoly bychom rádi poznamenali, že kanonický tvar DMRG vlnové funkce v rovnici (\ref{eq_mps_can}) odpovídá tzv. jednoorbitalovému DMRG algoritmu. Z důvodu výrazně lepší konvergence se prakticky výhradně využívá tzv. dvouorbitalový algoritmus, ve kterém tenzor $C$, jehož prvky se v dané iteraci optimalizují, zahrnuje na místo jednoho dva fyzikální indexy, tedy dva sousední orbitaly, např. $C^{n_p n_{p+1}}_{lr}$.

\section{Algoritmus renormalizační grupy matice hustoty}
\label{section_algorithm}

Mezi důležité vlastnosti DMRG vlnových funkcí patří nejen velmi kompaktní forma, ale také možnost efektivního výpočtu středních hodnot, např. energie. Jen díky tomu lze metodu DMRG využít pro variační výpočet energie. 
Skutečnost, že DMRG vlnovou funkci lze popsat pomocí malého množství parametrů totiž samo o sobě nezaručuje efektivní výpočet středních hodnot.

Efektivní výpočet středních hodnot si pro jednoduchost ukážeme na příkladu operátoru $\hat{O}$ působícího na dva sousední orbitaly, $l$ a $l + 1$. Takový lokální operátor lze rozepsat do báze odpovídající obsazenosti orbitalů $l$ a $l+1$\footnote{Na ostatní orbitaly působí jako identita.}

\begin{equation}
  \hat{O} = \sum_{ \{ n^{\prime} \}, \{n \} } O_{n_l^{\prime} n_{l+1}^{\prime}; n_l n_{l+1}} | n_l^{\prime} n_{l+1}^{\prime} \rangle \langle n_l n_{l+1} |,
  \label{eq_o}
\end{equation}

\noindent
kde $O_{n_l^{\prime} n_{l+1}^{\prime}; n_l n_{l+1}}$ jsou maticové elementy operátoru $\hat{O}$. 

\begin{figure}[!ht]
  \begin{center}
    \includegraphics[width=15cm]{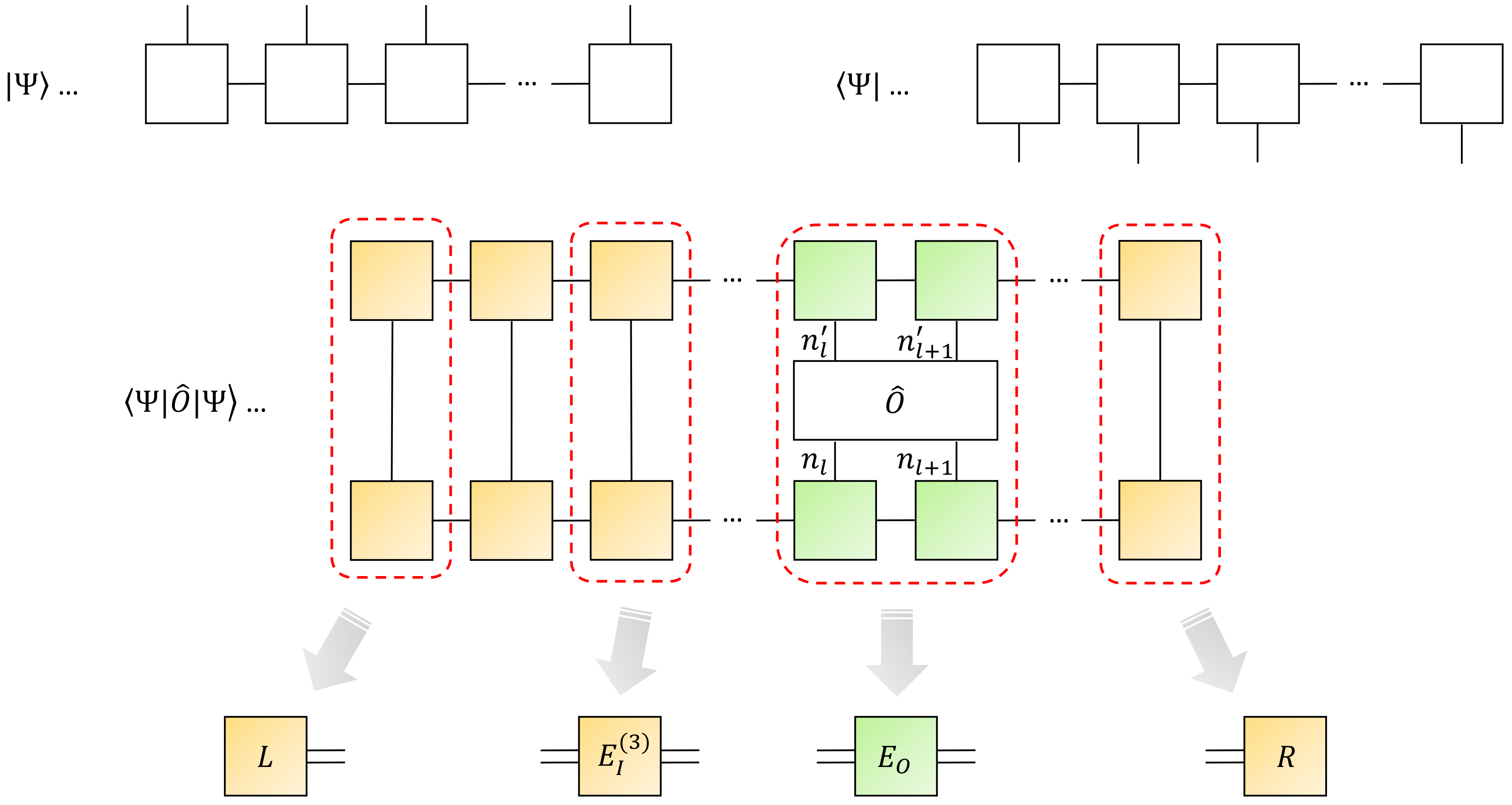}
  \end{center}
  \caption{Grafický zápis střední hodnoty lokálního oprátoru $\hat{O}$ působícího na orbitaly $l$ a $l+1$. Žlutě jsou označeny orbitaly, na které působí jako identita, zeleně pak orbitaly $l$ a $l+1$, na které působí netriviálně. Na obrázku jsou dále zobrazeny definice částečně zkontrahovaných tenzorů $L$, $E_I^{(i)}$, $E_O$ a $R$. Dvojitá čára odpovídá složenému indexu vzniklého z horního a spodního pomocného indexu, tzn. dimenze $M^2$.}
  \label{fig_exp_val}
\end{figure}

Nežli rozepisovat střední hodnotu operátoru $\hat{O}$ algebraicky, raději na obrázku \ref{fig_exp_val} představíme velmi intuitivní grafický zápis. Obdobně pomocným (horizontálním) indexům, kde přechod od řádkového ke sloupcovému vektoru znamenal přehození čáry indexu z pravé na levou stranu (viz. obrázek \ref{tensors}a), je přechod od normálního tvaru vlnové fukce (tzv. ket) k její Hermitovsky sdružené podobě (tzv. bra) charakterizován přehozením svislých čar fyzikálních indexů od shora dolů. Na obrázku \ref{fig_exp_val} je vidět, že oprátor $\hat{O}$ působí neidenticky pouze na orbitaly $l$ a $l+1$ a má 4 fyzikální indexy ($n_l, n_{l+1}, n_l^{\prime}, n_{l+1}^{\prime}$), což je v souladu s rovnicí (\ref{eq_o}).

Základem efektivního výpočtu střední hodnoty operátoru $\hat{O}$, je správné pořadí sumací (kontrakcí) tenzorové sítě z obrázku \ref{fig_exp_val}. Například naivní postup, při kterém bychom nejprve provedli sumaci přes pomocné indexy, by vedl zpět na $4^k$ složitost. Efektivního výpočtu můžeme dosáhnout, pokud provádíme kontrakci postupně z leva do prava. Na obrázku \ref{fig_exp_val} jsou znázorněny také částečně zkontrahované tenzory $L$, $E_I^{(i)}$, $E_O$ a $R$. S jejich pomocí lze střední hodnotu operátoru $\hat{O}$ vyjádřit následovně

\begin{equation}
  \langle \Psi | \hat{O} | \Psi \rangle = L E^{(2)}_I E^{(3)}_I \cdots E^{(l-1)}_I E_O E^{(l+2)}_I \cdots E^{(n-1)}_I R .
\end{equation}

\noindent
Jedná se o postupná násobení vektoru dimenze $M^2$ maticemi $M^2 \times M^2$. Protože je celkový počet orbitalů $k$, výpočetní náročnost střední hodnoty operátoru $\hat{O}$ může být shora omezena polynomiální funkcí $\mathcal{O}(kM^4)$. Tuto výpočetní náročnost lze při využití kanonického tvaru vlnové funkce (viz. sekce \ref{canonical_form}) a rozdělení kontrakcí pomocných indexů normální a Hermitovsky sdružené vlnové funkce dále snížit na $\mathcal{O}(kM^3)$.

Kvantově chemický Hamiltonián (\ref{H2q}) je součtem $\mathcal{O}(k^4)$ členů odpovídajících součinu molekulových integrálů s kreačními a anihilačními operátory, které působí maximálně na 4 orbitaly (bez ohledu na jejich celkový počet).
Fakt, že se obecně nejedná o sousední orbitaly, nepřináší žádné principiální komplikace\footnote{Kontrakce se provede v zásadě stejným způsobem, jen s tím, že předsumovaných tenzorů $E_O$ bude více (na místech, kde působí kreační a anihilační operátory). Pro zajištění správných antikomutačních relací kreačních a anihilačních operátorů se využívá tzv. Jordanovy-Wignerovy transformace \cite{jw}.}
 a mohlo by se zdát, že celková výpočetní náročnost kvantově chemické verze metody DMRG je rovna 
$\mathcal{O}(k^5M^3)$ ($k^4$-krát $kM^3$). Nicméně při využití efektivních předsumovaných operátorů, jejichž popis je nad rámec tohoto úvodního článku, lze výpočetní náročnost snížit na $\mathcal{O}(k^3M^3) + \mathcal{O}(k^4 M^2)$ \cite{white_1999}.

\subsection{Variační minimalizace energie}

\begin{figure}[!ht]
  \begin{minipage}{0.48\textwidth}
    \begin{center}
      \includegraphics[width=7.5cm]{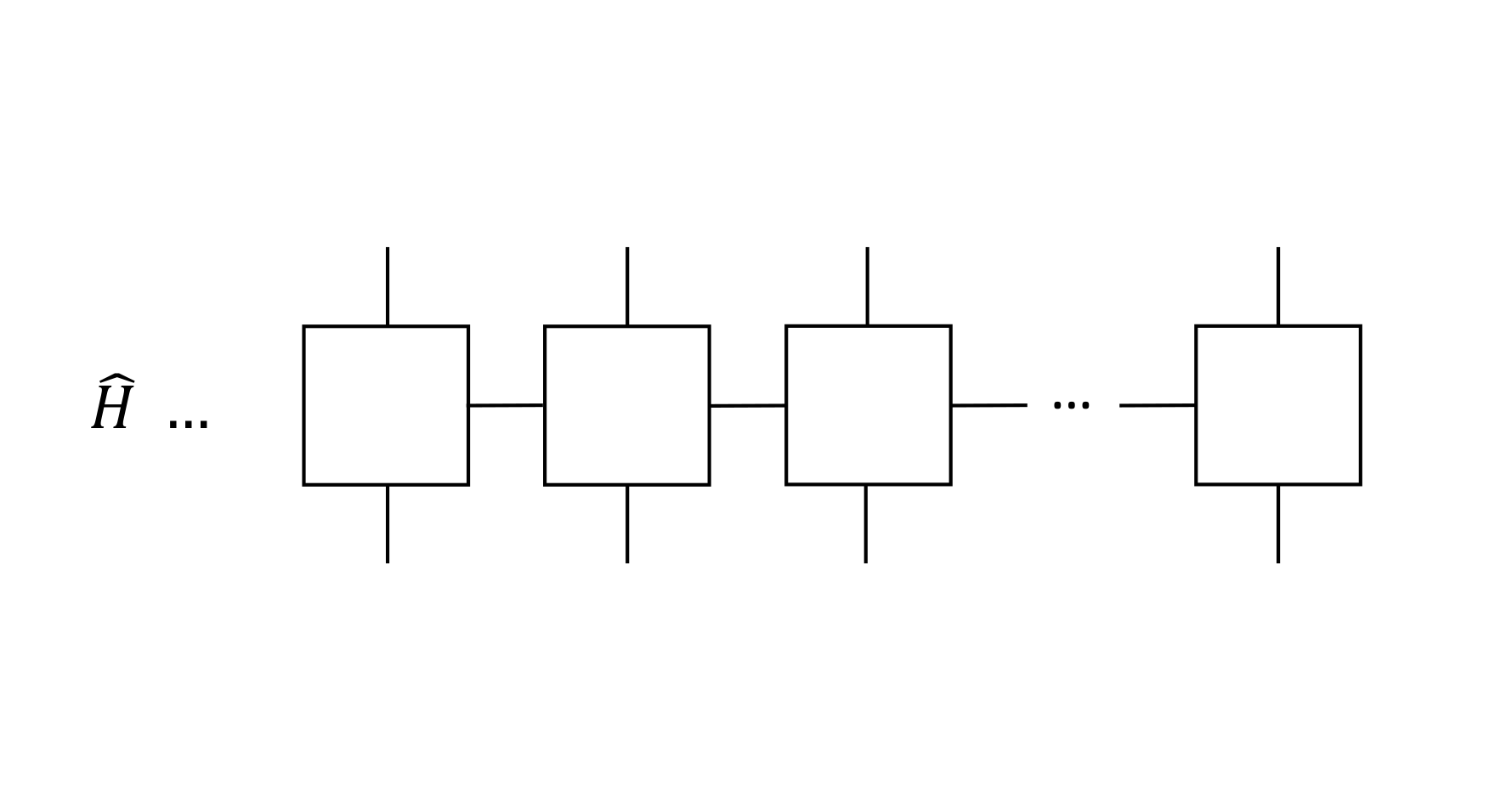} \\
      (a)
     \end{center}
  \end{minipage}
  \begin{minipage}{0.48\textwidth}
    \begin{center}
      \includegraphics[width=7.5cm]{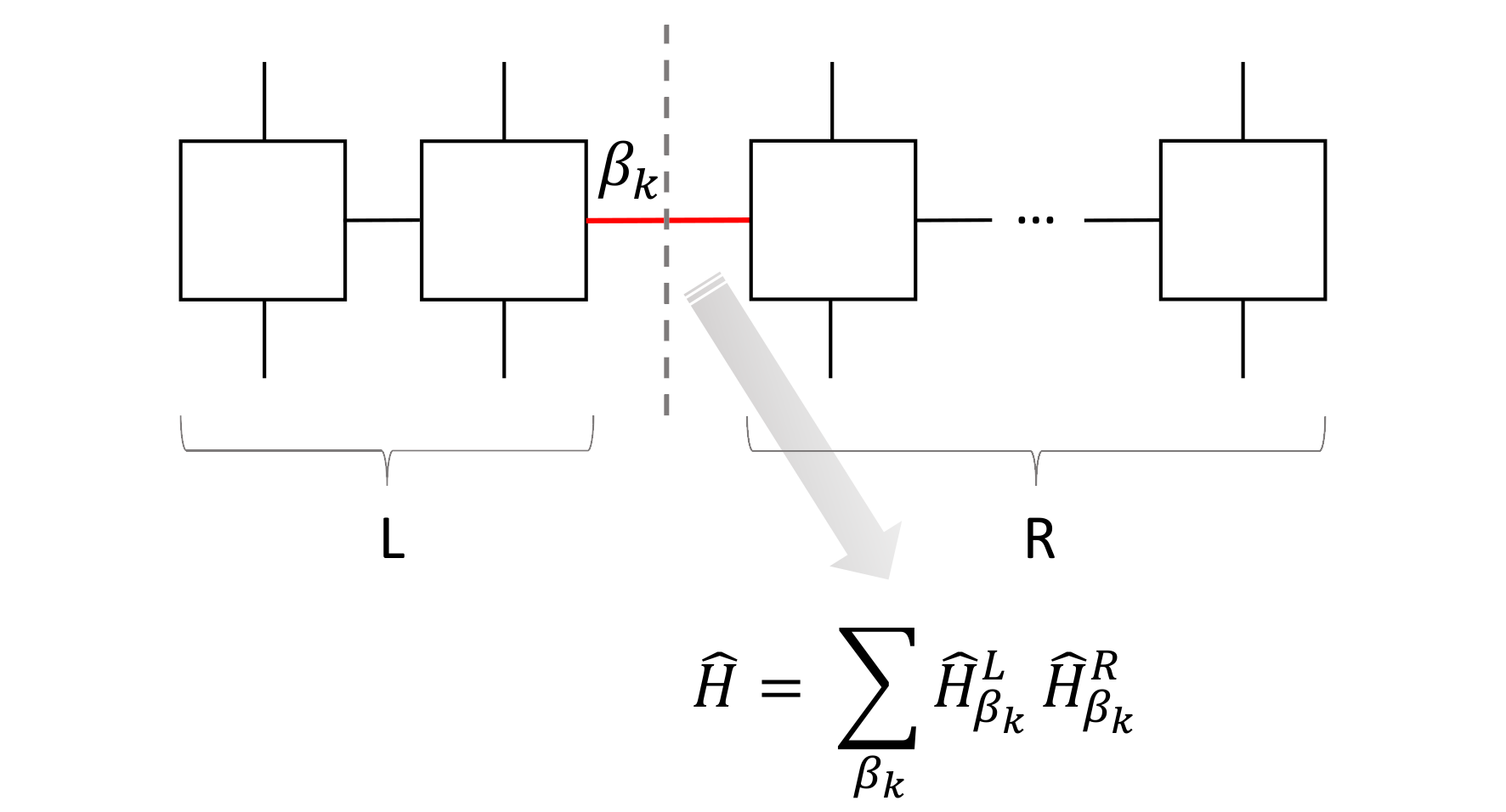} \\
      (b) 
    \end{center}
  \end{minipage}
  \vskip 1cm
  \begin{minipage}{0.48\textwidth}
    \begin{center}
      \includegraphics[width=7.5cm]{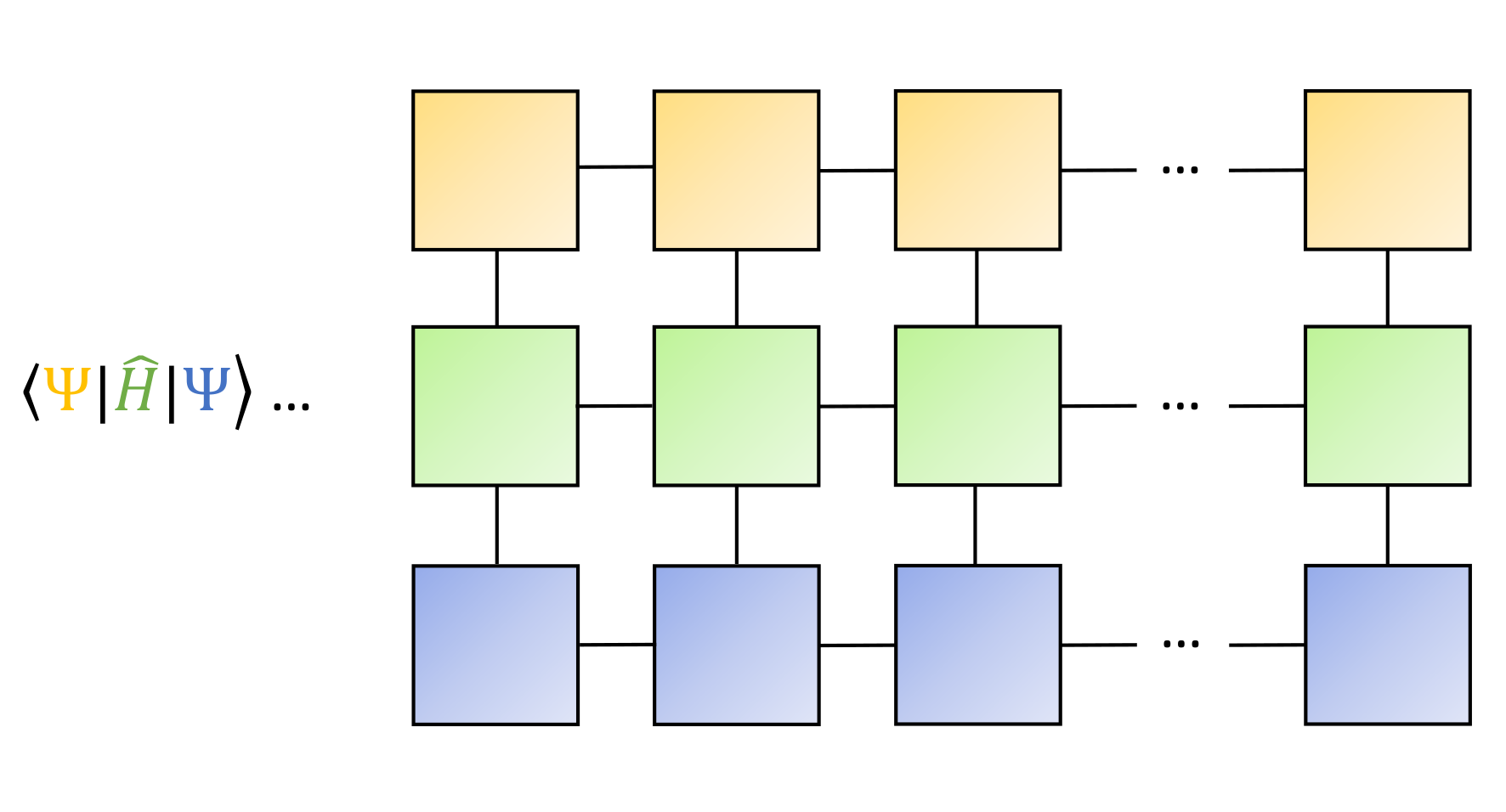} \\
      (c)
     \end{center}
  \end{minipage}
  \begin{minipage}{0.48\textwidth}
    \begin{center}
      \includegraphics[width=7.5cm]{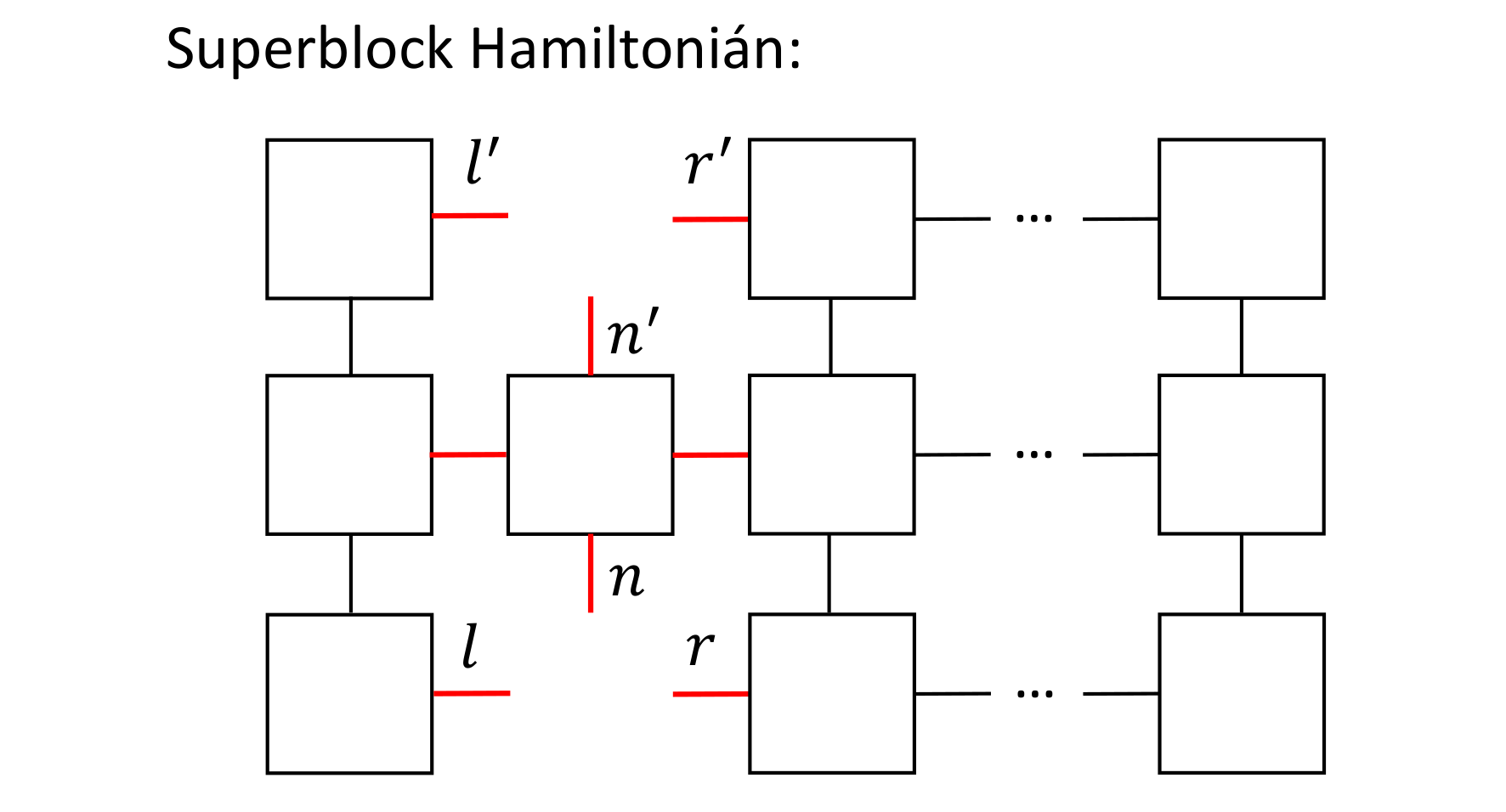} \\
      (d) 
    \end{center}
  \end{minipage}
  \caption{(a) Obecný tvar MPO Hamiltoniánu. (b) Pomocný (virtuální) index MPO Hamiltoniánu mající význam sumy interakčních členů levého a pravého bloku. (c) Střední hodnota Hamiltoniánu. (d) Superblock Hamiltonián, jehož diagonalizací se optimalizují prvky MPS matic příslušného orbitalu.}
  \label{fig_mpo}
\end{figure}

Obdobně, jako se dá
vlnová funkce libovolného stavu faktorizovat do MPS podoby \cite{vidal_2003}, dá se Hamiltonián (\ref{H2q}) faktorizovat do tzv. \textit{matrix product operator} (MPO) tvaru \cite{schollwock_2011}. Bohužel v tomto článku nemůžeme zabíhat do větších detailů. Jen zmiňme, že tak, jako má MPS vlnová funkce (vektor v Hilbertově prostoru) jednu sadu fyzikálních indexů odpovídající obsazenosti jednotlivých orbitalů, má MPO Hamiltonián (matice v Hilbertově prostoru), který je také graficky znázorněn na obrázku \ref{fig_mpo}a, dvě sady fyzikálních indexů. 

Pomocné (virtuální) indexy v případě MPO Hamiltoniánu reprezentují sumace přes interakční členy spojující orbitaly na levé a pravé straně od místa kontrakce \cite{chan_white_mpo}, viz. obrázek \ref{fig_mpo}b.

Zmíněná variační minimalizace energie, při které se optimalizují prvky MPS matic pouze jednoho orbitalu, ve skutečnosti odpovídá vlastnímu problému matice Hamiltoniánu, který se v tradiční formulaci založené na renormalizační grupě nazývá Hamiltonián superbloku \cite{schollweck_2005}. Tato matice vznikne částečnou kontrakcí tenzorové sítě střední hodnoty Hamiltoniánu tak, jak je znázorněno na obrázku \ref{fig_mpo}d\footnote{Z pohledu původní formulace renormalizační grupy odpovídá Hamiltonián superbloku maticové reprezentaci Hamiltoniánu v renormalizované mnoha-elektronové bázi levého a pravého bloku a daného orbitalu, viz. rovnice (\ref{eq_blocks}).}. 
Z důvodu velikosti této matice ($4M^2 \times 4M^2$) a faktu, že při typické aplikaci v kvantové chemii se zajímáme jen o několik málo nejníže ležících stavů, se tato matice obvykle diagonalizuje iterativně, např. Davidsonovým algoritmem (obdobně jako v CI).

\subsection{Výběr aktivního prostoru a řazení orbitalů}

Výběr správného aktivního prostoru je obecně netriviální záležitost, která v tradičním CAS\-SCF přístupu vyžaduje značnou chemickou intuici. Jak ale bylo nedávno ukázáno \cite{stein_2016, legeza_2003b, Boguslawski-2012b}, lze tuto činnost s pomocí nástrojů kvantové informatiky do velké míry zautomatizovat. Za tím účelem lze definovat tzv. entropie entanglementu \cite{eisert_tensor_networks}, které vyjadřují sílu korelace mezi částmi studovaného systému. Např. jednoorbitalová entropie $S_1^{(i)}$ vyjadřuje míru korelace daného ($i$-tého) orbitalu se všemi zbylými. Vyjadřuje tedy jistým způsobem důležitost daného orbitalu v aktivním prostoru podobně jako obsazovací číslo orbitalu. Dvouorbitalová entropie $S_2^{(ij)}$ pak kvantifikuje na míru korelace daného orbitalového páru ($ij$) se zbylými orbitaly. Důležitou veličinou je také tzv. vzájemná informace $I_{ij}$ definovaná následovně

\begin{equation}
  I_{ij} = S_1^{(i)} + S_1^{(j)} - S_2^{(ij)},
\end{equation}

\noindent
která vyjadřuje vzájemnou korelaci mezi dvěma orbitaly. Těchto veličin, které se dají v rámci přibližných DMRG výpočtů velmi snadno spočítat, se využívá v protokolu automatického výběru aktivního prostoru \cite{stein_2016}.

Jak bylo řečeno v kapitole \ref{section_wf}, DMRG vlnová funkce dokáže optimálně popsat jednodi\-men\-zi\-o\-nál\-ní systémy s krátkodosahovou interakcí, ve kterých jsou korelace mezi jednotlivými orbitaly velmi lokální. V případě obecných molekul s Coulombickou interakcí je situace o poznání složitější. Abychom docílili co nejlepší přesnosti, je potřeba studované molekuly co nejvíce přiblížit zmíněným jednodimenzionálním systémům, tzn. co nejvíce lokalizovat korelace mezi jednotlivými orbitaly. Pokud pracujeme s fixní bází molekulových orbitalů\footnote{Další možností je optimalizace báze orbitalů v průběhu DMRG algoritmu \cite{krumnow_2016}.}, ať už delokalizovaných, či lokálních, nástroj, kterým toho můžeme dosáhnout jsou permutace pořadí orbitalů v jednodimenzionálním (MPS a MPO) uspořádání.

Dnes už standardním postupem optimalizace pořadí orbitalů tak, aby meziorbitalové korelace byly co nejvíce lokální, je výpočet vzájemné informace $I_{ij}$ \cite{rissler_2006, barcza_2011} a nejčastěji minimalizace funkce 

\begin{equation}
  \text{cost}_2 = \sum_{ij} I_{ij} d_{ij}^2,
  \label{cost2}
\end{equation}

\noindent
kde $d_{ij}$ označuje vzdálenost orbitalů $i$ a $j$ v jednodimenzionálním uspořárání, např. $d_{ij} = 1$ pro sousední orbitaly. Vzájemnou informaci lze snadno spočítat z DMRG vlnové funkce \cite{barcza_2015}. 

Určení optimálního pořadí orbitalů je ve skutečnosti těžký, NP-úplný problém. Obecně však platí, že skutečně optimální pořadí není nezbytně nutné, stačí pořadí, které je rozumné. V praktickém protokolu optimalizace pořadí orbitalů se nejdříve z relativně levných výpočtů s malým $M$ určí přibližná vzájemná informace, $I_{ij}$, a s pomocí metod teorie grafů minimalizuje funkce z rovnice (\ref{cost2}) \cite{barcza_2011}. Základním pravidlem je, že při použití lokalizovaných orbitalů by měla vzdálenost $d_{ij}$ odrážet skutečnou vzdálenost orbitalů. Při použití delokalizovaných HF orbitalů je výhodné, aby vazebné a anti-vazebné orbitaly byly umístěny vedle sebe.

Pokud je metoda DMRG použita v aktivním prostoru, tedy ne v úplném orbitalovém prostoru, je možné optimalizavat tvar orbitalů obdobným způsobem, jako v případě metody CASSCF. Pak hovoříme o metodě DMRG-CASSCF, či DMRG-SCF \cite{zgid_2008_scf, ghosh_2008}. Její výhodou je možnost pracovat s výrazně většími aktivními prostory (až cca 40 orbitalů) než v případě metody CASSCF.

\section{Vlastnosti metody renormalizační grupy matice hustoty}
\label{properties}

Ještě nežli ilustrujeme metodu DMRG na několika konkrétních aplikacích, rádi bychom explicitně zdůraznili její tři formální vlastnosti, které jsou považovány v kvantové chemii za velmi důležité. 

\subsection{Variačnost}

Protože pracujeme s explicitním tvarem DMRG vlnové funkce (viz. kapitola \ref{section_wf}) a DMRG energie se počítá jako střední hodnota Hamiltoniánu, představuje tedy variační horní mez k přesné energii\footnote{Přesně variační je ve skutečnosti pouze jednoorbitalová verze metody DMRG, kterou jsme se výhradně zabývali. Dvouorbitalová verze může vykazovat malé odchylky od variačnosti.}. Při zvětšování dimenze MPS matic $M$ (zvyšování počtu variačních parametrů) konverguje DMRG energie shora k přesné energii.

\subsection{Multireferenčnost}

Při pohledu na tvar DMRG vlnové funkce (\ref{eq_dmrg_wf}) vidíme, že v zásadě nerozlišujeme mezi obsazenými a virtuálními orbitaly. Se všemi se pracuje stejným způsobem a již rozvoj (\ref{eq_fci}) napovídá, že HF referenční determinant nemá žádný speciální význam\footnote{Informace o HF referenci se v praxi pro urychlení konvergence využívá při počátečním odhadu MPS matic. Nicméně v principu by mělo jít zkonvergovat DMRG vlnovou funkci i s náhodným počátečním odhadem MPS matic.}. Z tohoto důvodu můžeme očekávat (a také pozorovat), že metoda DMRG bude velmi vhodná pro popis silné (statické) korelace v multireferenčních problémech. Naproti tomu popis dynamické korelace, pro který je informace o tom, které orbitaly jsou obsazené a které virtuální, velmi užitečná, zřejmě nebude optimální.

\subsection{Size konzistence}

Rozvoj DMRG vlnové funkce (\ref{eq_dmrg_wf}) je při použití lokalizované báze a správné volbě aktivního prostoru také tzv. size konzistentní. To znamená, že jestliže separujeme interagující systém $AB$ na dva neinteragující podsystémy $A$ a $B$, odpovídá celková vlnová funkce $\ket{\Psi_{AB}}$ součinu vlnových funkcí podsystémů $\ket{\Psi_A}$ a $\ket{\Psi_B}$ (má také MPS tvar) a celková enerige je součtem jejich energií. Podmínka lokalizované báze je potřeba, aby šly vlnové funkce neinteragujících podsystémů vyjádřit pomocí disjunktních množin orbitalů.

\section{Příklady použití metody renormalizační grupy matice hustoty v kvantové chemii}
\label{section_applications}

Od průkopnické práce Whita a Martina \cite{white_1999} s první aplikací metody DMRG v \textit{ab initio} kvantové chemii uběhlo již téměř dvacet let a během této doby byla kvantově chemická verze metody DMRG použita na celou řadu nejrůznějších molekulárních sysémů. 
V následující kapitole z důvodu rozsahu zmíníme jen některé charakteristické aplikace, velmi detailní výčet aplikací lze nalézt např. v nedávném přehledném článku \cite{wouters_review}.

První aplikace v kvantové chemii vedly přirozeně na malé molekuly \cite{chan_2003, legeza_2003a, legeza_2003c, chan_2004, chan_2004b, kurashige_2009}. Jednalo se o téměř přesná řešení Schr\"{o}dingerovy rovnice (v bázích o mnoho větších než umožňuje konvenční FCI). Za zmínku stojí např. studie molekuly vody v triple-zeta dvakrát polarizační bázi \cite{chan_2003}, ve které by přesný FCI rozvoj vyžadoval enormních $5.6 \times 10^{11}$ determinantů. Systematické DMRG výpočty se zvyšujícím se $M$ (řádově až k několika tisícům) ukázaly, že je snadné výsledky extrapolovat na $M = \infty$ a dosáhnout tak FCI energie.

Další studií, která z našeho pohledu stojí za zmínku, je práce týkající se zakázaného křížení molekuly LiF \cite{legeza_2003c}, neboť nevyužívá fixního $M$, ale metodu, která jej dokáže měnit tak, aby bylo dosaženo předem stanovené přesnosti, tzv. \textit{dynamical block state selection} \cite{legeza_2003a}.

Velmi důležitou prací je také výpočetní studie disociační křivky molekuly dusíku \cite{chan_2004b}. Disociace této molekuly je notoricky známý výpočetně složitý multireferenční problém (jedná se o přetržení trojné vazby). Ve zmíněné práci bylo ukázáno, že metoda DMRG je schopná popsat tento proces velmi vyváženě, s chybou, která prakticky nezávisí na mezijaderné vzdálenosti. Naproti tomu jednoreferenční CC metoda zahrnující excitace až šesti elektronů (UCCSDTQPH), přestože je schopná tuto molekulu popsat extrémně přesně pro rovnovážnou geometrii, vykazuje velmi rychlý nárůst chyby pro více disociovanou vazbu.

Typickou skupinou silně korelovaných molekul, pro které se metoda DMRG ukázala jako velmi vhodná, jsou molekuly komplexů přechodných kovů. 
Jejich elektronová struktura je komplikovaná energeticky blízko si ležícími \textit{d} (případně i \textit{f}) orbitaly.
V posledních letech vzniklo mnoho prací studující komplexy s jedním, dvěma, nebo doknce i více atomy přechodných kovů, například  \cite{marti_2008, kurashige_2009, amaya_2015, kurashige_2013, sharma_2014b, chalupsky_2014}. Společným poznatkem prvních studií na Cu$_2$O$_2$ modelech aktivního centra tyrosinázy \cite{marti_2008, kurashige_2009} je, že v případě elektronické struktury komplexů přechodných kovů lze získat zkovergované energetické rozdíly s daleko menším $M$, než jaké by vyžadovala zkonvergovaná absolutní energie.

Za zmínku také zcela jistě stojí nedávné studie výpočetně náročných a z chemického hlediska velmi důležitých komplexů přechodných kovů, konkrétně Mn$_4$Ca aktivní centrum fotosystému II\footnote{Zde byl použit aktivní prostor přesahující 50 orbitalů.} \cite{kurashige_2013} a také v bio-anorganické chemii doslova všudypřítomných [Fe-S] klastrů (konkrétně [2Fe-2S] a [4Fe-4S]) \cite{sharma_2014b}. 
O důležitosti této práce svědčí fakt, že [Fe-S] klastry tvoří důležité kofaktory nejrůznějších metaloenzymů a jsou zodpovědné za celou řadu chemických procesů, které se odehrávají v živých orgranismech (např. fotosyntéza, či dýchání)  \cite{beinert_1997, fontecave_2006}.

Poslední práci, kterou bychom rádi zmínili v souvislosti s malými molekulami je relativistická studie hydridu thalia (TlH) \cite{knecht_2014}. Důležitá je především z toho důvodu, že se jedná o čtyřsložkové výpočty (tedy plně relativistické). Jak bylo zmíněno v úvodní kapitole, také relativistické kvantově chemické problémy totiž mohou být vyjádřeny pomocí \mbox{Hamiltoniánu (\ref{H2q})}\footnote{Ve skutečnosti se nemusíme omezit pouze na elektronovou strukturu, pomocí Hamiltoniánu (\ref{H2q}) lze vyjádřit (a metodou DMRG počítat) také např. strukturu atomových jader \cite{legeza_2015}.}.

Další velkou skupinou sytémů, na kterou byla kvantově chemická verze metody DMRG velmi úspěšně aplikována jsou velké molekuly s jednodimenzionální topologií. Pro tyto systémy je metoda DMRG z důvodů diskutovaných v kaptiole \ref{section_wf} ideální a v těchto případech jsou aktivní prostory obsahující řádově 100 orbitalů dosažitelné.
Mezi příklady takových prací patří například studie radikálového charakteru acenů \cite{hachmann_2007}, magnetismu u oligofenylkarbenů \cite{yanai_2009}, či polarizabilit oligopolydiacetylenů \cite{dorando_2009}.

\subsection{Metody pro výpočet chybějící dynamické korelace}

Na kvantově chemickou verzi metody DMRG se dnes nejčastěji pohlíží jako na účinnou výpočetní metodu velmi dobře popisující především silnou korelaci a to v rámci nějakého relativně velkého aktivního prostoru.
Obvykle není ani možné, ani z důvodů zmíněných v kapitole \ref{properties} příliš výhodné, popisovat také dynamickou korelaci zahrnutím všech orbitalů do aktivního prostoru. 
Proto byly v předešlých letech vyvinuty techniky kombinující kvantově chemickou verzi metody DMRG s jinými metodami vhodnými pro popis chybějící dynamické korelace. 

Mezi ně patří například přístupy založené na poruchové teorii, konkrétně obdoba metody CASPT2 označovaná jako DMRG-CASPT2 \cite{Kurashige-2011}, či multireferenční poruchová teorie založená na MPS formulaci \cite{sharma_2014c}. Neporuchové přístupy zahrnují kombinaci metody DMRG s interně kontrahovanou MR CI \cite{saitow_2013}, či tzv. kanonickou transformaci vyvinutou Yanaiem a Chanem (\textit{canonical transformation}, CT) \cite{ctchan1,ctchan2,neuscamman_2010_irpc}, která ve skutečnosti odpovídá aproximativní interně kontrahované unitarní multireferenční CC metodě.

V naší skupině jsme nedávno vyvinuli alternativní přistup kombinující metody DMRG a CC, konkrétně tzv. \textit{tailored coupled clusters} \cite{kinoshita_2005} externě korigované pomocí MPS vlnové funkce \cite{veis_2016, veis_2018}.

\section{Závěr}

Kvantově chemická verze metody DMRG se v předešlých téměř dvaceti letech prokázala jako robustní a velmi úspěšná metoda vhodná pro popis elektronové struktury molekul obsahujích velký počet silně korelovaných elektronů. 

Z principiálních důvodů vyplývajících z obecného tvaru DMRG vlnové funkce je tato metoda vhodná především pro výpočty silně korelovaných molekul s lineární topologií, kde umožňuje výpočty s řádově 100 orbitaly v aktivním prostoru. Nicméně také v případě obecných silně korelovaných molekul umožňuje výpočty s výrazně vetšími aktivními prosory (až cca 50), než dovoluje např. metoda CASSCF. V tomto ohledu je velmi vhodná pro výpočty elektronové struktury komplexů přechodných kovů, především těch obsahujících více atomů přechodných kovů. Nedávné první aplikace na takové systémy se ukázaly jako velmi slibné \cite{kurashige_2013, sharma_2014b}.

Z chemického hlediska je velice důležité alespoň částečně zahrnout dynamickou korelaci, která v samotném DMRG popisu chybí. To je také v současnosti předmětem aktivního výzkumu. Především kombinace DMRG s vhodnými \textit{lineárně škálujícími} metodami pro výpočet zbylé dynamické korelace jsou podle našeho názoru velmi nadějné. Takové výpočetní metody v budoucnu zcela jistě umožní výpočty výrazně větších multireferenčních systémů (např. v bio-anorganické chemii).

\section*{Poděkování}

Rádi bychom poděkovali finanční podpoře z Grantové agentury České republiky (granty č. 18- 18940Y, 16-12052S, 18-24563S) a Ministerstva školství mládeže a tělovýchovy (grant č. LTAUSA17033).

Dále bychom rádi poděkovali Lukášovi Sobkovi za cenné připomínky k rukopisu tohoto článku.

\bibliography{reference}

\end{document}